\documentclass[twocolumn,showpacs,amsmath,amssymb,10pt]{revtex4}
\usepackage{amsfonts}
\usepackage{amssymb}
\usepackage{mathrsfs}
\usepackage{amsthm}
\usepackage{graphicx}
\usepackage{amsmath}
\newcommand{\cc}{\mathbb{C}}
\newcommand{\tr}{\mathrm{Tr}}
\newcommand{\hh}{\mathcal{H}}
\newcommand{\bb}{\mathcal{B}}

\newcommand{\psii}[1]{\ket{\psi_{#1}}}
\newcommand{\fif}[1]{\ket{\varphi_{#1}}}
\newcommand{\chir}[1]{\ket{\chi_{#1}}}
\newcommand{\fifr}[1]{\ket{\phi_{#1}}}
\newcommand{\1}{\ket{1}}
\newcommand{\2}{\ket{2}}
\newcommand{\3}{\ket{3}}
\newcommand{\4}{\ket{4}}
\newtheorem{thm}{Theorem}

\theoremstyle{definition}

\newcommand{\ot}[0]{\otimes}
\newcommand{\bei}{\begin{itemize}}
\newcommand{\eei}{\end{itemize}}
\newcommand{\ket}[1]{|#1\rangle}

\def\<{\langle}
\def\>{\rangle}
\newcommand{{\Cn}}{{\mathbb{C}^4}}
\newcommand{{\CN}}{{\mathbb{C}^{2n}}}
\newcommand{{\BC}}{{\mathcal{B}(\mathbb{C}^n)}}
\newcommand{{\BBC}}{{\mathcal{B}(\mathbb{C}^{2n})}}
\def\oper{{\mathchoice{\rm 1\mskip-4mu l}{\rm 1\mskip-4mu l}{\rm 1\mskip-4.5mu l}{\rm 1\mskip-5mu l}}}
\begin{document}

\title{\textbf{Indecomposable optimal entanglement witnesses in $\Cn \ot \Cn$}}

\author{Dariusz  Chru\'sci\'nski and Filip A. Wudarski\\
Institute of Physics, Nicolaus Copernicus University,\\
Grudzi\c{a}dzka 5/7, 87--100 Toru\'n, Poland}

\begin{abstract}
We provide two 1-parameter families of indecomposable entanglement witnesses in $\Cn \ot \Cn$.
Following recent paper by Ha and Kye [Phys. Rev. A {\bf 84}, 024302 (2011)]  we show that these EWs are optimal and hence provide the strongest tool in entanglement theory to discriminate between separable and entangled states. As a byproduct we show that these EWs detect quantum entanglement within a family of generalized Horodecki states.
\end{abstract}

\maketitle

The most general approach to discriminate between separable and entangled states of a quantum composite system living in $\mathcal{H}_A \ot \mathcal{H}_B$ is  based on the notion of positive maps or equivalently
 entanglement witnesses (EWs) \cite{HHHH,Terhal1,Guhne}.  A state $\rho$ in $\mathcal{H}_A \ot \mathcal{H}_B$ is separable iff $(\oper_A \ot \Phi)\rho \geq 0$ for all
linear positive maps $\Phi : \mathcal{B}(\mathcal{H}_B) \rightarrow \mathcal{B}(\mathcal{H}_A)$. A hermitian operator $W\in\bb(\hh_A\otimes\hh_B)$ is an entanglement witness iff:
i) it is not positively defined, i.e. $W \ngeqslant 0$, and ii) $\tr(W\sigma)\ge0$ for all separable states $\sigma$. Furthermore,
a bipartite state $\rho$ living in $\hh_A\otimes\hh_B$ is entangled iff there exists an EW $W$ detecting this state, i.e.
such that $\tr(W\rho)<0$. Due to the well known duality between linear  maps and operators in $\bb(\hh_A\otimes\hh_B)$ these two
approaches are fully equivalent. A positive map $\Phi$ is optimal if $\Phi - \Lambda^{\rm CP}$ is not a positive map for an arbitrary completely positive map $\Lambda^{\rm CP}$. In terms of EWs optimality has more operational meaning: given an EW $W$ let $D_W$ denote a subset of states detected by $W$. Following \cite{Lew} an entanglement witness $W$ is optimal if there is no EW $W'$ such that $D_W \subset D_{W'}$. It is, therefore, clear that the knowledge of optimal EWs is sufficient for the full characterization of separable/entangled states.

Recall, that an EW $W$ is decomposable if $W = P + Q^\Gamma$, where $P,Q \geq 0$ and $Q^\Gamma$ denotes the partial transposition. It is clear from the definition that a decomposable EW cannot detect an entangled PPT state ($\rho$ is PPT if its partial transposition $\rho^\Gamma \geq 0$). Equivalently, a linear positive map $\Phi$ is decomposable if $\Phi = \Lambda_1 + \Lambda_t \circ T$, where $\Lambda_1,\Lambda_2$ are completely positive and $T$ denotes transposition. An indecomposable positive map $\Phi$ is called nd-optimal if $\Phi - \Lambda_{\rm D}$ is not a positive map for an arbitrary decomposable positive map $\Lambda_{\rm D}$. Similarly, given an indecomposable EW $W$ let $\widetilde{D}_W$ denote a subset of PPT states detected by $W$. $W$ is nd-optimal if there is no indecomposable EW $W'$ such that $\widetilde{D}_W \subset \widetilde{D}_{W'}$. It shows that nd-optimal EWs {\em optimally} detect PPT entangled states. One has \cite{Lew} the following

\begin{thm}
An entanglement witness $W$ in nd-optimal if and only if both $W$ and $W^\Gamma$ are optimal.
\end{thm}

Unfortunately, in spite of the considerable
effort (see e.g. \cite{Choi}--\cite{Filip-II}), the structure of indecomposable positive maps (equivalently, indecomposable entanglement witnesses) is rather poorly understood.

In a recent paper \cite{Kye-PRA} Ha and Kye proved the nd-optimality of 1-parameter family of maps defined as follows:
for $a,b,c \geq 0$ we define a map $\Phi[a,b,c] : M_3 \rightarrow M_3$ by the following formula
%\begin{widetext}
\begin{equation*}\label{}
    \Phi[a,b,c](X)  = \frac{1}{a+b+c} \left( \begin{array}{cccc} y_1 & -x_{12} & -x_{13}  \\
    - x_{21} & y_2 & -x_{23}  \\
    - x_{31} & - x_{32} & y_3
     \end{array} \right) \ ,
\end{equation*}
where $x_{ij}$ are matrix elements of $X \in M_3$, and
\begin{equation*}
    y_i = \sum_{j=1}^3 \, A_{ij}[a,b,c]\, x_{jj}\ ,
\end{equation*}
with $A_{ij}$ being the matrix elements of the following circulant doubly stochastic matrix
\begin{equation*}
    A[a,b,c] =  \frac{1}{a+b+c}\left(  \begin{array}{ccc} a & b & c  \\ c& a & b  \\ b & c & a  \end{array} \right)\ .
\end{equation*}
It is well known that $\Phi[a,b,c]$ is a positive map in $M_3$ if and only if
\begin{equation}\label{P-3}
 a+b+c\geq 2\ , \ \ a\leq 1 \Rightarrow bc \geq (1-a)^2\ .
\end{equation}
Moreover, $\Phi[a,b,c]$ is positive but not completely positive if in addition to (\ref{P-3}) one has $a < 2$.
In this case the corresponding entanglement witness reads as follows
\begin{align}
 & W[a,b,c] = \sum_{i=1}^3 \Big[\, a\, |ii\>\<ii| + b\, |i,i+1\>\<i,i+1| \nonumber \\ & + c\, |i,i+2\>\<i,i+2|\, \Big]
   - \sum_{i\neq j}  |ii\>\<jj|\ .
\end{align}
In a recent paper \cite{FilipI} we analyzed a subclass defined by
\begin{equation}\label{elipsa}
a\leq 1\ , \ \    a+b+c=2\ , \ \ bc=(1-a)^2\ .
\end{equation}
Note, that the corresponding $W[a,b,c]$ is decomposable if and only if $b=c$. Ha and Kye \cite{Kye-PRA} proved the following
\begin{thm}
If $a,b,c$ satisfy (\ref{elipsa}), then $W[a,b,c]$ define an nd-optimal EW.
\end{thm}

The aim of the present paper is to generalize this result for maps in $M_4$.
Therefore, for $a,b,c,d \geq 0$ satisfying
\begin{equation}\label{}
    a+b+c+d=3\ ,
\end{equation}
let us define a map $\Phi[a,b,c,d] : M_4 \rightarrow M_4$ by the following formula
%\begin{widetext}
\begin{equation*}\label{}
    \Phi[a,b,c,d](X)  = \frac 13  \left( \begin{array}{cccc} y_1 & -x_{12} & -x_{13} & - x_{14} \\
    - x_{21} & y_2 & -x_{23} & -x_{24} \\
    - x_{31} & - x_{32} & y_3 & -x_{34} \\
    - x_{41} & -x_{42} & - x_{43} & y_4 \end{array} \right)\ ,
\end{equation*}
%\end{widetext}
where
\begin{equation*}
    y_i = \sum_{j=1}^4 \, A_{ij}[a,b,c,d]\, x_{jj}\ ,
\end{equation*}
with $A_{ij}[a,b,c,d]$ being the matrix elements  of the following doubly stochastic circulant matrix
\begin{equation*}
    A[a,b,c,d] = \frac 13 \left(  \begin{array}{cccc} a & b & c & d \\ d& a & b &  c \\ c & d & a & b \\ b & c & d & a \end{array} \right)\ .
\end{equation*}
It turns out \cite{Filip-II} that there are two classes of positive maps $\Phi[a,b,c,d]$ characterized by the following conditions:

$$ \mbox{Class I} $$
\begin{equation}\label{I}
    a+c=2\ , \ \ \ b+d =1\ ,\ \ \  bd=(1-a)^2\ .
\end{equation}

$$ \mbox{Class II} $$
\begin{equation}\label{II}
  a+c=1\ , \ \ \ b+d =2\ , \ \ \  ac=(1-b)^2\ .
\end{equation}
Note, that maps $\Phi[1,1,1,0]$ and $\Phi[0,1,1,1]$ are already known in the literature. The former belonging to Class I is a generalized Choi map 
%constructed by Osaka \cite{Osaka} 
whereas the latter belonging to Class II is the standard reduction map
$\, \Phi[0,1,1,1](X) = \frac 13(\mathbb{I}_4 {\rm Tr}\, X - X)$.

The  entanglement witness $W[a,b,c,d]$ corresponding to $\Phi[a,b,c,d]$ reads as follows
%\begin{widetext}
\begin{align*}
 & W[a,b,c,d] = \sum_{i=1}^4 \Big[\, a\, |ii\>\<ii| + b\, |i,i+1\>\<i,i+1| + \\
 &  c\, |i,i+2\>\<i,i+2| + d\, |i,i+3\>\<i,i+3|\, \Big]
   - \sum_{i\neq j}  |ii\>\<jj|\ ,
\end{align*}
%\end{widetext}
where we add modulo 4. It turns out \cite{Filip-II} that $W[a,b,c,d]$ is indecomposable iff $b\neq d$.

\begin{thm}
If $a,b,c,d \geq 0$ satisfy (\ref{I})  or (\ref{II}) and $a < 1$, then $W[a,b,c,d]$ defines an optimal entanglement witness.
\end{thm}

{\bf Proof}: let us recall \cite{Lew} that to prove optimality of $W$ living in $\mathcal{H}_A \ot \mathcal{H}_B$ it is sufficient to construct  a family of product vectors $\psi \ot \phi \in \mathcal{H}_A \ot \mathcal{H}_B$ satisfying $\< \psi \ot \phi|W|\psi \ot \phi \>=0$ such that vectors $\psi \ot \phi$ span $\mathcal{H}_A \ot \mathcal{H}_B$. We consider separately both classes.

\noindent 1) Class I together with $a<1$: let
\begin{equation}\label{tt1}
    t = \frac{1-a}{b}\ , \ \   \ t_1 = \frac{2-a}{1+c}\ .
\end{equation}
Note that for $a<1$ one has $b>0$ and hence the formula for $t$ is well defined.
Let us define the following 16 vectors $\psi_k \ot \varphi_k$: for $k=1,\ldots,4$
\begin{equation*}\label{}
    |\psi_k\> = \sum_{l=1}^4 e^{i\lambda_{kl}} |l\> \ , \ \ |\varphi_k\> = |\psi_k^*\>\ ,
\end{equation*}
with random phases $\lambda_{kl}$. Moreover
\begin{align*}
&\psii{5}=\sqrt{t}\1+\2 \ ,  & \quad&   \fif{5}=\sqrt{t}\1+t\2\ , \\
&\psii{6}=\sqrt{t}\1+i\2 \ , & \quad& \fif{6}=\sqrt{t}\1-it\2\ , \\
&\psii{7}=\sqrt{t}\2+\3 \ , & \quad&  \fif{7}=\sqrt{t}\2+t\3\ , \\
&\psii{8}=\sqrt{t}\2+i\3 \ , & \quad&  \fif{8}=\sqrt{t}\2-it\3\ , \\
&\psii{9}=\sqrt{t}\3+\4 \ , & \quad&  \fif{9}=\sqrt{t}\3+t\4\ , \\
&\psii{10}=\sqrt{t}\3+i\4 \ , & \quad&  \fif{10}=\sqrt{t}\3-it\4\ , \\
&\psii{11}=\sqrt{t}\4+\1 \ , & \quad& \fif{11}=\sqrt{t}\4+t\1\ , \\
&\psii{12}=\sqrt{t}\4+i\1\ , & \quad& \fif{12}=\sqrt{t}\4-it\1\ ,
\end{align*}
Finally, the remaining 4 vectors reads as follows
\begin{eqnarray*}
\psii{13}&=&\sqrt{t_1}\1+\sqrt{t_1}\2+\3+\4 \ , \\
\fif{13}&=&\sqrt{t_1}\1+\sqrt{t_1}\2+t_1\3+t_1\4\ , \\
\psii{14}&=&\sqrt{t_1}\1+i\2+i\3+\sqrt{t_1}\4 \ , \\
\fif{14}&=&\sqrt{t_1}\1-it_1\2-it_1\3+\sqrt{t_1}\4\ ,\\
\psii{15}&=&\sqrt{t_1}\1+\sqrt{t_1}\2+i\3+i\4 \ , \\
\fif{15}&=&\sqrt{t_1}\1+\sqrt{t_1}\2-it_1\3-it_1\4\ , \nonumber\\
\psii{16}&=&i\1+\sqrt{t_1}\2+\sqrt{t_1}\3+i\4 \ , \\
\fif{16}&=&-it_1\1+\sqrt{t_1}\2+\sqrt{t_1}\3-it_1\4\ .
\end{eqnarray*}
Assuming (\ref{I}) one easily proves that for $k=1,\ldots,16$
\begin{equation*}\label{}
    \< \psi_k \ot \varphi_k|W[a,b,c,d]|\psi_k \ot \varphi_k\> = 0 \ ,
\end{equation*}
and that both sets of vectors $\{ \psi_k \ot \varphi_k \}$ and $\{ \psi_k \ot \varphi_k^*\}$ span $\cc^4 \ot \cc^4$.

\noindent 2) Class II together with $a>0$: let us define the following 16 vectors $\chi_k \ot \phi_k$: for $k=1,\ldots,5$
\begin{equation*}\label{}
    |\chi_k\> = \sum_{l=1}^4 e^{i\nu_{kl}} |l\> \ , \ \ |\phi_k\> = |\chi_k^*\> \ ,
\end{equation*}
with random phases $\nu_{kl}$. Moreover
\begin{align}
&\chir{6}=\sqrt{t}\1+\3 \ , & \quad& \fifr{6}=\sqrt{t}\1+t\3 \ ,\nonumber\\
&\chir{7}=\sqrt{t}\1+i\3\ ,& \quad&\fifr{7}=\sqrt{t}\1-it\3\ ,\nonumber\\
&\chir{8}=\sqrt{t}\2+\4\ , & \quad&\fifr{8}=\sqrt{t}\2+t\4\ ,\nonumber\\
&\chir{9}=\sqrt{t}\2+i\4\ ,& \quad&\fifr{9}=\sqrt{t}\2-it\4\ ,\nonumber\\
&\chir{10}=\sqrt{t}\3+\1\ ,& \quad&\fifr{10}=\sqrt{t}\3+t\1\ ,\nonumber\\
&\chir{11}=\sqrt{t_2}\3+i\1\ ,& \quad&\fifr{11}=-i\sqrt{t_2}\3+t_2\1\ ,\nonumber\\
&\chir{12}=\sqrt{t}\4+\2\ ,& \quad&\fifr{12}=\sqrt{t}\4+t\2\ ,\nonumber\\
&\chir{13}=\sqrt{t_2}\4+i\2\ ,& \quad&\fifr{13}=-i\sqrt{t_2}\4+t_2\2\ ,\nonumber
\end{align}
together with the remaining 3 vectors
\begin{eqnarray*}
\chir{14}&=&\sqrt{t_3}\1+i\2+i\3+\sqrt{t_3}\4 \ , \\
\fifr{14}&=&\sqrt{t_3}\1-it_3\2-it_3\3+\sqrt{t_3}\4\ ,\\
\chir{15}&=&\sqrt{t_3}\1+\sqrt{t_3}\2+i\3+i\4 \ , \\
\fifr{15}&=&\sqrt{t_3}\1+\sqrt{t_3}\2-it_3\3-it_3\4\ , \nonumber\\
\chir{16}&=&i\1+\sqrt{t_3}\2+\sqrt{t_3}\3+i\4 \ , \\
\fifr{16}&=&-it_3\1+\sqrt{t_3}\2+\sqrt{t_3}\3-it_3\4\ ,
\end{eqnarray*}
where
\begin{equation}\label{t2t3}
 t_2 = \frac{1+a}{b} \ , \ \ \ t_3= \frac{5-2a}{1+2c}\ ,
\end{equation}
and $t$ is defined in (\ref{tt1}).
Note that for $a,b,c,d$ satisfying (\ref{II}) one has $b > 0$ and hence both $t$ and $t_2$ are  well defined.
Assuming (\ref{II}) one easily proves that for $k=1,\ldots,16$
\begin{equation*}\label{}
    \< \chi_k \ot \phi_k|W[a,b,c,d]|\chi_k \ot \phi_k\> = 0 \ ,
\end{equation*}
and that both sets of vectors $\{ \chi_k \ot \phi_k \}$ and $\{ \chi_k \ot \phi_k^*\}$ span $\cc^4 \ot \cc^4$. \hfill $\Box$

Let us observe that the above constructions of product vectors break up if $a=1$. Indeed, if $a=1$ then $t=0$ and the corresponding vectors are no longer linearly independent. Therefore, both Choi maps $\Phi[1,1,1,0]$ and $\Phi[1,0,1,1]$ (Class I) and decomposable map $\Phi[1,1,0,1]$ are excluded. Note, that if  $|\psi\> = \sum_{k=1}^k e^{i\lambda_k} |k\>$, then
$$ \< \psi \ot \psi^*|W[a, b, c, d]|\psi \ot \psi^*\> = 0 \ , $$
for arbitrary phases $\lambda_k$. Vectors $\psi \ot \psi^*$ span 13-
dimensional subspace in $\cc^4 \ot \cc^4$. Interestingly, vectors
$\psi \ot \psi$ span only 10-dimensional subspace. It
turns out that for EWs arising form generalized Choi
maps, i.e. $W[1, 1, 1, 0]$ and $W[1, 0, 1, 1]$, there are only
13 linearly independent product vectors satisfying $\< \psi\ot \phi|W|\psi \ot \phi\>=0$.
The same situation we already met for the original Choi maps in $M_3$. In this case we have only 7
vectors (9 is sufficient for optimality). However, Choi maps in $M_3$ are known to be extremal and hence they are optimal as well. We conjecture that the same applies here, that is, both $W[1, 1, 1, 0]$ and $W[1, 0, 1, 1]$ are extremal EWs without the spanning property.

Note, that optimal entanglement witnesses $W[a,b,c]$ detect entanglement within well known Horodecki states \label{Hor-3}
\begin{equation}\label{}
    \rho^{(3)}_\alpha = \frac 17 \Big(2 P^+_3 + \alpha \Pi_1 + (5-\alpha)\Pi_2 \Big) \ ,
\end{equation}
where $P^+_3$ denotes the maximally entangled state in $\cc^3 \ot \cc^3$ and $\Pi_k = \frac 13 \sum_{l=1}^3 |l,l+k\>\<l,l+k|$. It is well known that $\rho^{(3)}_\alpha$ defines a legitimate state iff $\alpha \in [0,5]$. Moreover, it is PPT entangled iff $\alpha \in [1,2) \cup (3,4]$.
One may easily check that for any such $\alpha$ one can always find $a,b,c$ satisfying (\ref{elipsa}) such that ${\rm Tr}(\rho^{(3)}_\alpha W[a,b,c]) <0$. Interestingly, the above family of states may be easily generalized for $n>3$ (cf. \cite{HOR}). If $n=4$ one finds
\begin{equation}\label{}
    \rho^{(4)}_\alpha = \frac{1}{16} \Big(3 P^+_4 + \alpha \Pi_1 + 3 \Pi_2 + (10-\alpha)\Pi_3\Big) \ ,
\end{equation}
where now $\Pi_k = \frac 14 \sum_{l=1}^4 |l,l+k\>\<l,l+k|$. It was shown \cite{HOR} that $\rho^{(4)}_\alpha$ is PPT entangled iff $\alpha \in [1,3) \cup (7,9]$. Again, simple calculation shows that for any such $\alpha$ one can always find $a,b,c,d$ satisfying (\ref{I}) or (\ref{II}) such that ${\rm Tr}(\rho^{(4)}_\alpha W[a,b,c,d]) <0$.

In conclusion, we proved that two classes of EWs described by (\ref{I}) and (\ref{II}) provide nd-optimal EWs.
It would be interesting to know if the map ƒ$\Phi[a, b, c,d]$  generates
an extremal ray of the cone of all positive linear maps whenever the conditions (\ref{I}) or (\ref{II}) hold.
As a byproduct we show that these EWs detect quantum entanglement within a family of generalized Horodecki states.

\end{document}